\documentclass[12pt,twoside]{article}
\usepackage[mathscr]{eucal}
\usepackage{amsmath,amsfonts,amssymb,amsthm}
\usepackage[matrix,arrow]{xy}
\usepackage{times}
\usepackage{srcltx}
\usepackage{float} 
\usepackage{hyperref}
\usepackage{graphicx}
\usepackage{epstopdf}
\usepackage{multibox}
\usepackage{dcolumn}

\def\d_Vphi{\text{d}_V\hspace{-0.06em}\phi}
\def\d_Vphibar{\text{d}_V\hspace{-0.06em}\bar\phi}
\def\d_Vxi{\text{d}_V\hspace{-0.06em}\xi}

\def\AKSZ{Alexandrov:1997kv}
\def\BGST{Barnich:2004cr}

\def\be{\begin{eqnarray}}
\def\ee{\end{eqnarray}}
\def\beann{\begin{eqnarray*}}
\def\eeann{\end{eqnarray*}}
\def\beq{\begin{equation}}
\def\eeq{\end{equation}}
\def\ba{\begin{array}}
\def\ea{\end{array}}
\def\ben{\begin{enumerate}}
\def\een{\end{enumerate}}
\def\bea{\begin{eqnarray}}
\def\eea{\end{eqnarray}}

\def\5{\bar }
\def\6{\partial }
\def\7{\hat }
\def\4{\tilde }
\advance\textheight\topskip
\renewcommand{\tilde}{\widetilde}
\renewcommand{\hat}{\widehat}
\newtheorem{prop}{Proposition}[section]

\renewcommand{\simeq}{\cong}

\newcommand{\bref}[1]{\textbf{\ref{#1}}}

\renewcommand{\dim}{\mathop{\mathrm{dim}}}
\renewcommand{\mod}{\mathop{\mathrm{mod}}}

\newcommand{\p}[1]{|#1|}
\newcommand{\gh}[1]{\mathrm{gh}(#1)}

\newcommand{\dd}{\partial}
\renewcommand{\d}{\partial}

\renewcommand{\geq}{\,{\geqslant}\,}
\renewcommand{\leq}{\,{\leqslant}\,}

\newcommand{\binner}[2]{%
  {\langle}\kern-4.15pt{\langle}#1{,}\,#2{\rangle}\kern-4.15pt{\rangle}}

\newcommand{\pb}[2]{\left\{{}#1{},{}#2{}\right\}}

\newcommand{\half}{\mathchoice{%
    \ffrac{1}{2}}{\frac{1}{2}}{\frac{1}{2}}{\frac{1}{2}}}

\newcommand{\ffrac}[2]{\raisebox{.5pt}%
  {\footnotesize$\displaystyle\frac{#1}{#2}$}\kern1pt}

\newcommand{\dl}[1]{\mathchoice{\ffrac{\dd}{\dd #1}}{\frac{\dd}{\dd
      #1}}{\ffrac{\dd}{\dd #1}}{\ffrac{\dd}{\dd #1}}}

\newcommand{\dover}[2]{\ffrac{\dd #1}{\dd #2}}

\newcommand{\ddl}[2]{\ffrac{\dd #1}{\dd #2}}

\newcommand{\vdl}[1]{\ffrac{{\delta}}{\delta #1}}

\newcommand{\vddr}[2]{\ffrac{\delta^R #1}{\delta #2}}
\newcommand{\vddl}[2]{{\ffrac{\delta #1}{\delta #2}}}

\newcommand{\derham}{\boldsymbol{d}}

\newcommand{\manifold}[1]{\mathscr{#1}}
\newcommand{\manX}{\manifold{X}}

\newcommand{\manM}{\manifold{M}}
\newcommand{\manN}{\manifold{N}}

\def\cA{\mathcal{A}}

\def\cC{\mathcal{C}}

\def\cF{\mathcal{F}}
\def\cG{\mathcal{G}}

\def\cI{\mathcal{I}}

\def\cN{\mathcal{N}}

\numberwithin{equation}{section} \makeatletter
\@addtoreset{equation}{section}

\begin{document}

\def\mytitle{A Poincar\'e lemma for sigma models of
  AKSZ type}

\pagestyle{myheadings}
\markboth{\textsc{\small Barnich, Grigoriev}}{%
  \textsc{\small Local BRST cohomology for AKSZ-type sigma models}}
\addtolength{\headsep}{4pt}

\begin{flushright}\small
ULB-TH/09-08, Preprint ESI 2141 (2009)
\end{flushright}

\begin{centering}

  \vspace{1cm}

  \textbf{\Large{\mytitle}}

  \vspace{1.5cm}

  {\large Glenn Barnich$^{a,*}$ and Maxim Grigoriev$^{b,a}$}

\vspace{.5cm}

\begin{minipage}{.9\textwidth}\small \it \begin{center}
   $^{a}$Physique Th\'eorique et Math\'ematique, Universit\'e Libre de
   Bruxelles\\ and \\ International Solvay Institutes, \\ Campus
   Plaine C.P. 231, B-1050 Bruxelles, Belgium \end{center}
\end{minipage}

\vspace{.5cm}

\begin{minipage}{.9\textwidth}\small \it \begin{center}
   $^{b}$Tamm Theory Department, Lebedev Physics
   Institute,\\ Leninsky prospect 53, 119991 Moscow, Russia\end{center}
\end{minipage}

\end{centering}

\vspace{1cm}

\begin{center}
  \begin{minipage}{.9\textwidth}
    \textsc{Abstract}. For a sigma model of AKSZ-type, we show that
    the local BRST cohomology is isomorphic to the cohomology of the
    target space differential when restricted to coordinate
    neighborhoods both in the base and in the target. An analogous
    result is shown to hold for the cohomology in the space of
    functional multivectors. Applications of these latter cohomology
    classes in the context of the inverse problem of the calculus of
    variation for general gauge systems are also discussed.
  \end{minipage}
\end{center}


\vfill

\noindent
\mbox{}
\raisebox{-3\baselineskip}{%
  \parbox{\textwidth}{\mbox{}\hrulefill\\[-4pt]}}
{\scriptsize$^*$Research Director of the Fund for
  Scientific Research-FNRS (Belgium).}

\thispagestyle{empty}
\newpage

\begin{small}
{\addtolength{\parskip}{-1.5pt}
 \tableofcontents}
\end{small}
\newpage

\section{Introduction}
\label{sec:introduction}

The Batalin-Vilkovisky formalism has originally been devised as a
means to control gauge symmetries during perturbative quantization of
systems with a complicated gauge algebra
\cite{Batalin:1981jr,Batalin:1983wj,Batalin:1983jr,Batalin:1984ss,%
  Batalin:1985qj} (see e.g.~\cite{Henneaux:1992ig,Gomis:1995he} for
reviews). In this context, some questions of physical interest, such
as the classification of divergences or anomalies arising during
renormalization can be efficiently reformulated in terms of ``local
BRST cohomology'', i.e., the cohomology groups of the
antifield-dependent BRST differential in the space of local
functionals\footnote{With perturbative quantum field theory in mind,
  one might be tempted to use the gauge fixed, on-shell nilpotent BRST
  differential. Why it is far more transparent to fix the gauge
  through a canonical transformation while keeping the antifields
  instead of reducing to a Lagrangian submanifold when taking locality
  into account is explained in \cite{Barnich:1999cy,Barnich:2003tr}.}
(see e.g.~\cite{Piguet:1995er,Barnich:2000zw} for reviews).  On the
classical level, these groups control the deformation theory for gauge
systems and encode generalized global symmetries and conservations
laws.

Even though the BV master action is usually constructed using an
algorithmic procedure with input a classical action and a generating
set of gauge symmetries, it is sometimes more natural to define a
theory directly in terms of a master action. This is the case for
instance for gauge field theories associated to BRST first quantized
systems (see e.g.~\cite{Barnich:2003wj} for a review), such as open
string field theory or higher spin gauge fields
\cite{Thorn:1989hm,Ouvry:1986dv,Bengtsson:1986ys,Henneaux:1987cp}.

Another class of models that falls into this category are AKSZ sigma
models~\cite{Alexandrov:1997kv}, for which the master action on the
space of maps is directly constructed out of the geometrical data of
the base and target manifolds. In the AKSZ case, the target space is a
$QP$ manifold, a supermanifold equipped with a graded symplectic
structure and a compatible homological vector field. We will restrict
ourselves here to the case where the base space is $\Pi T\manX_0$, the
tangent space to a manifold $\manX_0$ with shifted parity of the
fibers equipped with the de Rham differential.

In the context of massless higher spin gauge fields, and in particular
in the so-called unfolded formulation
\cite{Vasiliev:1988sa,Vasiliev:1990en,Vasiliev:1992av,Vasiliev:2003ev},
the focus is in a first stage on the equations of motion, whether they
derive from an action principle or not. When translated in BRST
language, this amounts to defining a theory through a differential
which is not necessarily generated through the adjoint action of a
master action in an appropriate antibracket. For the AKSZ
construction, this means that one is mainly interested in the $Q$
structure and forgets about the $P$ structure. Such non-Lagrangian
AKSZ-type sigma models are directly related to a BRST extended version
of the non-linear unfolded formalism
\cite{Barnich:2005ru,Vasiliev:2005zu}.

In the same way as in the applications to soliton equations or quantum
field theory, algebraic control on the space of maps for sigma models
can be achieved in the context of the formal variational calculus,
where derivatives of fields are considered as independent coordinates
on so-called jet-spaces and local functionals are quotients of
horizontal $n$-forms modulo exact ones (see
e.g.~\cite{Andersonbook,Anderson1991,Dickey:1991xa,Olver:1993} for reviews).

The purpose of this paper is to show that, in coordinate
neighborhoods, the local BRST cohomology for Lagrangian and
non-Lagrangian AKSZ-type sigma models, and thus also for BRST extended
unfolded models, is isomorphic to the $Q$-cohomology in target space.

This result is then extended to the cohomology computed in the
space of functional multivectors.  These cohomology groups become
important in the non-Lagrangian setting.  For instance, they control
consistent deformations and global symmetries in this context. We also
show that the cohomology for higher multivectors controls weak Poisson
structures and their counterparts in the Lagrangian formalism known as
Lagrange structures~\cite{Lyakhovich:2004xd,Kazinski:2005eb}.  They
are therefore relevant for the inverse problem of the calculus of
variations applied to gauge systems.

The drastic simplification of the field theoretic cohomology is not
really surprising in view of the structure of the BRST differential
for AKSZ sigma models. Nevertheless, the precise isomorphism that we
have established gives concrete meaning to the notion of background
independence for non-Lagrangian AKSZ-type sigma models.

More interesting is the global situation with non trivial topology. In
a global approach, it is known how the cohomology of the bundle of
target over base space is reflected in the cohomology of the
variational bicomplex~\cite{Andersonbook}. What one then needs to
analyze is how this latter cohomology affects the so-called descent
equations that are used to compute the BRST cohomology in the space of
local functionals from the cohomology in the space of horizontal
forms. An elementary example of this interplay has been given in the
context of Einstein gravity, where the target space topology is non
trivial due to the determinant condition on the metric
\cite{Barnich:1995ap}. When taking into account in addition the
topology of the base space, the problem becomes more involved and the
right spectral sequence for the computation needs to be
identified. The appropriate framework to address this global question
is likely to be the $\cC$-spectral sequence by Vinogradov (see
\cite{Vinogradov:1977,Vinogradov:1978,Vinogradov:1984,vinogradov:2001}
and references therein). We plan to return to the global question
elsewhere.

Let us end this introduction by briefly reviewing related
literature. 

The first AKSZ sigma model for which local BRST cohomology
has been explicitly computed and shown to reduce to a cohomology
problem in target space is Chern-Simons \cite{Delduc:1990je,%
  Blasi:1990xz,Lucchesi:1991pb} (see also
\cite{Barnich:1994db,Barnich:2000zw}). 

For general AKSZ sigma models, we have followed in our paper the
general strategy proposed for BF theory in
\cite{Lucchesi:1992gp,Piguet:1992yg} and reviewed in
\cite{Piguet:1995er}. The proof in these papers is, however,
incomplete as the contractible pairs have not been correctly
identified. The correct identification has been discussed in details
using Young diagrams in \cite{Henneaux:1998rp}, although in a slightly
different context: these authors considered the gauge part of models
involving form-fields, whereas here we need to apply their results to
the space-time part of the antifield-dependent BRST differential.

In the non-linear unfolded off-shell formalism, it has been shown in
\cite{Vasiliev:2005zu} that the $Q$-cohomology in target space gives
rise to interesting field theoretic invariants like actions and
conserved charges (see also \cite{Barnich:2005ru}).  From this
perspective, what we have shown here is in some sense the inverse
statement: all field theoretic invariants which can be represented as
BRST cohomology classes in the space of local functionals in the
fields and their space-time derivatives arise from $Q$-cohomology in
the case of AKSZ type sigma models.

While completing this paper, we came across reference
\cite{Bonechi:2009kx}, where a reduction of the BV formalism for AKSZ
sigma models to base cohomology is discussed along different
lines. The assumptions underlying our main result amount to
trivializing the base cohomology. Our result can then be understood as
a concrete proof that, for AKSZ sigma models, such a reduction does
indeed occur in cohomology, or in other words, that there is a
quasi-isomorphism between the classical field theoretic BV formalism
and the classical BV formalism in target space.  From a technical
point of view, this concrete proof is possible because in the
jet-space approach, there is precise algebraic control over the space
of maps.

As said above, it would be interesting to extend this concrete proof
to the case of non trivial topology. Concerning the quantum BV
formalism, it is clear from renormalization theory that the coupling
constants need to play an active role in a precise, non formal,
definition of the field theoretic $\Delta$-operator. How this can be
done on the level of cohomology is discussed in
\cite{Barnich:1998qz,Barnich:2000me}.

\section{AKSZ construction}

\subsection{ BRST differential}
\label{sec:non-lagrangian-aksz}

{\em The construction of the field theoretic BRST differential on the
  space of maps from differentials in base and target space is briefly
  recalled. }

Consider two $Q$ manifolds, i.e., supermanifolds equipped with an odd
nilpotent vector field~\cite{Schwarz:1992gs}.  The first, called the
base manifold, is denoted by $\manX$. It is equipped with a grading
$\mathrm{gh}_\manX$ and its odd nilpotent vector field is denoted by
$\derham,\,\mathrm{gh}_\manX(\derham)=1$. Furthermore, the existence
of a volume form $d\mu$ preserved by $\derham$ is also assumed. As
implied by the notation, the basic example for $\manX$ is the odd
tangent bundle $\Pi T\manX_0$ to some manifold $\manX_0$ which has a
natural volume form and is equipped with the de Rham differential. We
restrict ourselves to this case below. If $x^\mu$ and $\theta^\mu$ are
coordinates on $\manX$ and the fibres of $\Pi T\manX_0$ respectively,
the differential and the volume form are given explicitly by
\begin{equation}
 \derham=\theta^\mu \dl{x^\mu}  \,,\quad d\mu=dx^0\dots dx^{n-1}
 d\theta^{n-1}\dots d\theta^0\equiv d^nx d^n\theta\,, \qquad
n=\dim \manX_0\,.
\end{equation}
The second supermanifold, called the target manifold, is denoted by
$\manM$ and equipped with another degree $\mathrm{gh}_\manM$. The odd
nilpotent vector field is denoted by $Q$ and
$\mathrm{gh}_{\manM}(Q)=1$.

Consider then the manifold of maps from $\manX$ to
$\manM$.\footnote{More generally, one could of course consider the
  space of sections of a bundle over $\manX$ with fibers isomorphic to
  $\manM$.} This space is naturally equipped with the total degree
$\gh{A}={\rm gh}_{\manM}(A)+ {\rm gh}_{\manX}(A)$ and an odd nilpotent
vector field $s$, $\gh s=1$. Using local coordinates
$x^\mu,\theta^\mu$ on $\manX$ and $\Psi^A$ on $\manM$ the expression
for $s$ is given by
\begin{equation}
\label{eq:s-AKSZ}
s=\int_{\manX}d^nx d^n\theta \left[\derham\Psi^A(x,\theta)+Q^A(\Psi(x,\theta))
\right]
\vdl{\Psi^A(x,\theta)}\,.
\end{equation}

Vector field $s$ can be considered as the BRST differential of a field
theory on $\manX_0$ and the construction described above is the
non-Lagrangian part of the AKSZ approach.  Indeed, the field space,
BRST differential, and ghost grading determine a gauge field theory
for which the physical fields can be identified with those carrying
ghost number zero, while the equations of motion, gauge symmetries,
and higher structures of the gauge algebra are encoded in the BRST
differential $s$\footnote{See e.g.~\cite{Henneaux:1992ig} for a review
  and~\cite{\BGST} for further developments in the non Lagrangian
  context.}.

\subsection{Bracket and BV master action}
\label{sec:qp-manifolds}

{\em The construction of the field theoretic bracket on the space of
  maps from a target space (odd) Poisson bracket is recalled.}

In the case where the target manifold $\manM$ is in addition equipped
with a compatible (odd) Poisson bracket $\pb{\cdot}{\cdot}_{\manM}$
and $Q=\pb{S}{\cdot}_\manM$ is generated by a master function $S$,
i.e., a function satisfying the classical master equation $\half
\pb{S}{S}_\manM=0$, one can construct a functional $\mathbf{S}$ on the
space of maps that can be interpreted either as the BV master action
or the BRST charge of the BFV Hamiltonian approach\footnote{The latter
  identification for an odd $\mathbf{S}$ of ghost number $1$ was
  proposed in~\cite{Grigoriev:1999qz}. We follow the conventions from
  this reference for brackets and functional derivatives.} of the
field theory on $\manX_0$.

More precisely, let $E^{AB}=\pb{\Psi^A}{\Psi^B}_\manM$ be an (odd)
Poisson bivector for the bracket $\pb{\cdot}{\cdot}_{\manM}$. The
associated bracket on the space of maps is given by
\begin{equation}
\label{eq:path-bracket}
\pb{F}{G}=(-1)^{(\p{F}+n)n}
\int d^nx d^n\theta\,
\big(\vddr{F}{\Psi^A(x,\theta)} E^{AB}(\Psi(x,\theta))
\vddl{G}{\Psi^B(x,\theta)}\big)\,.
\end{equation}
Here $F=F[\Psi],G=G[\Psi]$ are functionals on the space of maps.  If
the bracket on $\manM$ carries Grassmann parity $\kappa$ and ghost
number $k$, parity and ghost number of the functional Poisson bracket
are given respectively by $\kappa+n \mod 2$ and $k+n$. There is a
natural map from target space functions to functionals on the space of
maps: given a target space function $f$ one defines
\begin{equation}
\label{fF-map}
 \cI (f)=\int_{\manX} \Psi^* f=\int d^nx
 d^n\theta \, f(\Psi(x,\theta))\,,
\end{equation}
where $\Psi^*f$ is the pull-back of $f$ on $\manM$ to $\manX$ by the
map $\Psi$. The map $\cI$ is compatible with the differentials in the
sense that
\begin{equation}
 \cI(Qf)=s\cI(f)\,,
\end{equation}
when the map $\Psi$ is of compact support.  Moreover, $\cI$ is a
homomorphism of graded Lie superalgebras
\begin{equation}
\cI(\pb{f}{g}_\manM)= \pb{\cI(f)}{\cI(g)}\,,
\end{equation}
provided one shifts by $n=\dim{\manX_0}$ the ghost number and the
Grassmann parity for functions on $\manM$ in order to make $\cI$
compatible with the gradings.

If in addition the bracket $\pb{\cdot}{\cdot}_\manM$ is non
degenerate, i.e.,~ if $\manM$ is equipped with a symplectic
structure $E_{AB}$ determined by $E_{AB}E^{BC}=\delta_A^C$ and a
symplectic potential $V_A$ can be defined through $E_{AB} = (\partial_A
V_B - (-1)^{\p{A}\p{B}}
\partial_B V_A)(-1)^{\p{B}(\p{E}+1)}$, the functional vector field
induced by $\derham$ is Hamiltonian. Combining this with the
Hamiltonian induced by $S$, one obtains the functional,
\begin{gather}
  \mathbf{S}[\Psi]=
\int d^nx d^n \theta  \,\Big[\big(\derham \Psi^A(x,\theta)\big)
  V_A(\Psi(x,\theta))+S\big(\Psi(x,\theta)\big) \Big]\,,\\
  \half\pb{\mathbf{S}}{\mathbf{S}}=0\,,
\end{gather}
so that $s=\pb{\mathbf{S}}{\cdot}$.  Parity and ghost number of
$\mathbf{S}$ are $\p{\mathbf{S}}=\p{S}-n \mod 2$ and
$\gh{\mathbf{S}}=\gh{S}-n$. In particular, if
$\p{\mathbf{S}}=\gh{\mathbf{S}}=0$, functional $\mathbf{S}$ is to be
interpreted as a BV master action, while if
$\p{\mathbf{S}}=\gh{\mathbf{S}}=1$, functional $\mathbf{S}$ becomes
the BRST charge of a field theory on $\manX_0$.

This approach was originally proposed in~\cite{Alexandrov:1997kv} as a
method for constructing the BV formulation of topological sigma
models. Further developments can be found
in~\cite{Cattaneo:1999fm,Grigoriev:1999qz,Batalin:2001fc,Batalin:2001fh,%
  Cattaneo:2001ys,Park:2000au,Roytenberg:2002nu,Ikeda:2006wd} and
references therein.

\subsection{Examples}

{\em Some standard and not so standard examples of AKSZ sigma models
  are briefly reviewed.}

\subsubsection*{Chern-Simons theory}

The first example of an AKSZ sigma model discussed in \cite{\AKSZ} is
Chern-Simons theory. It corresponds to taking $\manM=\Pi \cG$ where
$\cG$ is a Lie algebra equipped with an invariant non degenerate
metric $g_{ab}$. This metric determines a non degenerate Poisson
structure on $\Pi \cG$. The AKSZ construction then gives the standard
BV master action for the Chern-Simons theory provided one takes
$\manX_0$ to be a 3-dimensional manifold. 

Note that the BRST differential is well defined for $\manX_0$ of any
dimension and does not require an invariant bilinear form. Such a BRST
differential describes the zero-curvature equations for a
$\cG$-connection and its natural gauge symmetry.

\subsubsection*{Poisson sigma model}

The Poisson sigma model \cite{Ikeda:1993fh,Schaller:1994es} can also
be formulated in the AKSZ framework~\cite{Cattaneo:1999fm}.  As a
target space, one takes $\manM=\Pi T^*\manN$, with $\manN$ a Poisson
manifold.  If $X^i,C_i$ are local coordinates on $\manM$, the $QP$
structure is determined by
\begin{equation*}
  \pb{X^i}{C_j}_{\manM}=\delta^i_j\,, \qquad Q=\pb{S}{\cdot}_\manM\,, 
\quad S=\half  C_i\alpha^{ij}(X)C_j\,,
\end{equation*}
where $\alpha^{ij}\d_i\wedge \d_j$ is a Poisson bivector.  The
homological vector field $Q$ defines the Poisson cohomology on
$\manM$.  As a spacetime, one takes a 2 dimensional manifold
$\manX_0$. The associated AKSZ master action is then the standard BV
master action for the Poisson sigma model.

\subsubsection*{BF theories}

For BF theories, the base space is $\manX=\Pi T\manX_0$ where
$\manX_0$ is an $n$-dimensional manifold. The target space is
$\manM=\Pi T^*(\Pi\cG)$ for even $n$ and $\manM=T^*(\Pi\cG)$ for odd
$n$ with its canonical odd (even) symplectic structure.  Using the
standard coordinates $c^a,b_a$ on $\manM$, with
$\gh{c^a}=1,\gh{b_a}=n-2$, the $QP$ structure is determined by
\begin{equation}
\pb{b_a}{c^b}_\manM=\delta_a^b\,,\qquad Q=\pb{\half b_a
  f^a_{bc}c^bc^c}{\cdot}_\manM\,. 
\end{equation}
Note that the bracket carries ghost number $1-n$ so that it induces
the standard BV antibracket on the space of maps. That the associated
AKSZ master action is indeed the standard master action for
non-abelian BF theory follows in particular from the fact that in
ghost number zero, the field content consists of $1$-forms and
$n-2$-forms.

\subsubsection*{Hamiltonian BFV systems with vanishing Hamiltonian}

Let us take as $\manM$ the extended phase space of the Hamiltonian BFV
formulation of a first class constrained system~\cite{Fradkin:1975cq,%
  Batalin:1977pb,Fradkin:1978xi} (see also
\cite{Henneaux:1985kr}). Such a system is described by a phase space
$\manM$, with coordinates $\Psi^A$ and a symplectic structure with
potential $V_A$, an associated non degenerate Poisson structure
$\pb{\cdot}{\cdot}_\manM$, a BRST charge $\Omega$ and a BRST invariant
Hamiltonian $H$. The associated BV formulation is governed by a master
action that can be directly constructed out of $\Omega$ and
$H$~\cite{Batalin88,Siegel:1989nh,Batlle:1989if,Fisch:1989rm}. It was
shown in~\cite{Grigoriev:1999qz} that, in the case of vanishing
Hamiltonian $H$, it is an AKSZ sigma model with target space the
symplectic manifold $\manM$, target space differential generated by
the BRST charge, $Q=\pb{\Omega}{\cdot}$ and base space $\manX_0$ a
``time'' line. The master action can be written as
\begin{equation}
  \mathbf{S}=\int\, dt d\theta
  \Big[\big(\derham\Psi^A(t,\theta)\big)V_A(\Psi(t,\theta))
+\Omega\big(\Psi(t,\theta)\big)\Big]\,,
\end{equation}
where $\derham=\theta\dl{t}$. From this point of view, a general AKSZ
sigma model appears simply as a multi-dimensional generalization of
this example.

The original BFV formulation has been constructed with quantization in
mind. Another class of AKSZ-type sigma models can be associated with
such quantum systems.  More precisely, the target space $Q$-structure
is determined by the BRST operator and the operator superalgebra of
the quantum constrained system.  The typical example is given by
higher spin fields as background fields for a quantized scalar
particle~\cite{Grigoriev:2006tt}.

\section{Generalities on jet-spaces and local BRST cohomology}
\label{sec:problem}

\subsection{Horizontal complex}
\label{sec:horizontal-complex}

{\em The definition of local functions, the horizontal complex and of local
functionals are recalled.}

Consider a graded vector space $F$ with coordinates $z^\alpha$,
$\alpha=1,\dots,m$. They include both ``fields'' and antifields. The
$\mathbb{Z}$ grading is denoted by $\mathrm{gh}$ (``ghost
number''). For simplicity we assume here that the Grassmann parity,
denoted by $\p{\cdot}$ is just $\mathrm{gh}$ modulo $2$. Consider
further the space $\manX_0\simeq \mathbb{R}^n$ (``spacetime'') with
coordinates $x^\mu$, and the jet-bundle associated to $F\times
\manX_0\stackrel{\pi}{\rightarrow} \manX_0$, with coordinates
$x^\mu,z^\alpha_{(\mu)}$. Local functions are functions of $x^\mu$ and
$z^\alpha_{(\mu)}$ that depend on the derivatives $z^\alpha_{(\mu)}$
up to some finite order\footnote{We follow the conventions of
  \cite{Andersonbook} for multi-indices and their summation. A summary
  can be found for instance in Appendix A of \cite{Barnich:2001jy}.},
where $\gh{x^\mu}=0$, $\gh{z^\alpha_{(\mu)}}=\gh{z^\alpha}$ and
$\p{z^\alpha_{(\mu)}}=\p{z^\alpha}$. The complex $\Omega^{*,*}$ of
horizontal forms $\omega=\omega(x,dx,[z])$ involves forms in $dx^\mu$
with coefficients that are local functions. As by now standard, we
identify $dx^\mu$ with $\theta^\mu$ which are taken to anticommute
with the odd elements among $z^\alpha_{(\mu)}$. The horizontal
differential is $d_H=\theta^\mu\partial_{\mu}$ where the total
derivative is defined by
\begin{eqnarray}
\partial_\mu=\ddl{}{x^\mu}+z^\alpha_{\mu}\ddl{}{z^\alpha}+\dots=
\ddl{}{x^\mu}+z^\alpha_{\mu(\nu)}\ddl{^S}{z^\alpha_{(\nu)}}.\label{eq:14} 
\end{eqnarray}

We assume that horizontal forms can be decomposed into field/antifield
independent and dependent parts,
$\omega=\omega(x,\theta,0)+\hat\omega(x,\theta,[z])$. The complex
involving the latter is denoted by $\hat \Omega^{*,*}$, where the
first degree refers to the ghost number while the second to the form
degree. A standard result (see
e.g.~\cite{Andersonbook,Dickey:1991xa,Olver:1993}) is that the
cohomology of this complex is trivial in form degrees less than $n$,
\begin{eqnarray}
H^k(d_H,\hat\Omega)=0,\ {\rm for}\ 0\leq k
<n.\label{eq:2}
\end{eqnarray}
The space of local functionals $\hat \cF^*$ is then defined as the quotient
space $\hat\Omega^{*,n}/\ d_H \hat\Omega^{*,n-1}$. The projection from
a representative $\omega^{g,n}\in \hat\Omega^{g,n}$ to an element of
the quotient space is often denoted by the integral sign, 
\begin{equation}
\hat
\cF^{g}\ni [\omega^{g,n}]=\int \omega^{g,n}\label{eq:18}.
\end{equation}
Euler Lagrange derivatives\footnote{Unless otherwise
  specified, all derivatives are left derivatives.} are defined by 
\begin{equation}
  \label{eq:19}
  \vddl{\omega^{g.n}}{z^\alpha}=\dover{\omega^{g,n}}{z^\alpha}-
    \d_\mu\dover{\omega^{g,n}}{z^\alpha_\mu}+\dots=(-)^{|\mu|}
\d_{(\mu)}\dover{^S\omega^{g,n}}{z^\alpha_{(\mu)}}.  
\end{equation}
A crucial property is that
\begin{equation}
\int \omega^{g,n}=0\iff
\vddl{\omega^{g.n}}{z^\alpha}=0\label{eq:26}.
\end{equation}

\subsection{BRST differential}
\label{sec:brst-diff-cohom}

{\em The definition of the field theoretic BRST differential is given.}

The BRST differential $s$ is an odd, nilpotent
evolutionary\footnote{See
  e.g.~\cite{Andersonbook,Dickey:1991xa,Olver:1993} for detailed
  discussions of vector fields on jet-bundles.} vector field, i.e.,~a
vector field of the form
\begin{eqnarray}
  \label{eq:13}
  s= \partial_{(\mu)} S^\alpha\ddl{^S}{z^\alpha_{(\mu)}}
\end{eqnarray}
with $S^\alpha$ local functions and $\gh s=1$, $s^2=0$. It follows
that $[s,\d_\mu]=0=[s,d_H]$, where the bracket denotes the graded
commutator. For later purposes, note that an evolutionary vector field
is entirely defined through its action on the undifferentiated fields,
$s z^\alpha= S^\alpha$, and the requirement that it commutes with the
total derivative $\d_\mu$.

\subsection{Local BRST cohomology}
\label{sec:local-brst-cohom}

{\em The definition of BRST cohomology in the space of local
  functionals is given and standard ways to compute it are sketched.}

Several cohomology groups can then be considered. For instance, the
cohomology of $s$ in the space of local functions or in the space of
horizontal forms. As mentioned in the introduction, especially
interesting in view of applications in classical and quantum
Lagrangian gauge field theories are the so-called local BRST
cohomology groups, i.e.,~the cohomology of $s$ in the space of local
functionals, $H^*(s,\hat\cF)$. By definition of local functionals,
$H^*(s,\hat\cF)\simeq H^{*,n}(s|d_H,\hat\Omega)$. The latter group is
defined by
\begin{equation}
  \label{eq:1}
  s \omega^{g,n}+d_H \omega^{g+1,n-1}=0,\quad \omega^{g,n}\sim
  \omega^{g,n}+ s \eta^{g-1,n}+d_H \eta^{g,n-1},
\end{equation}
with $\omega,\eta\in \hat\Omega$.  Using \eqref{eq:2}, one then finds
that $H^{g,n}(s|d_H,\hat\Omega)\simeq H^{g+n}(\tilde s,\hat\Omega)$
where $\tilde s=s+d_H$ and the grading is the sum of the ghost number
and the form degree. This statement summarizes the content of the
so-called ``descent equations'' which provide a standard way to
compute $H^*(s,\hat\cF)$ out of $H^*(s,\hat\Omega)$ (see
e.g.~\cite{Dubois-Violette:1985jb,Barnich:2000zw}).

\subsection{BRST cohomology for functional multivectors}
\label{sec:brst-cohom-funct}

{\em It is shown how the introduction of canonical momenta allows one
  to generalize local BRST cohomology to functional multivector
  fields.}

Another cohomology group that is usually considered is the commutator
cohomology of $s$ in the space of evolutionary vector fields. The
space of evolutionary vector fields is known to be isomorphic to the
space of functional univectors (see e.g.~\cite{Olver:1993}). More
generally, one can then consider the cohomology of $s$ in the space of
graded symmetric or skew-symmetric functional multivectors. Graded
skew-symmetric functional multivectors equipped with a functional
version of the Schouten-Nijenhuis bracket (also called BV antibracket)
are well known and extensively used for studying the Hamiltonian
structures of evolution equations. In the graded symmetric case, the
bracket is a functional version of the canonical graded Poisson
bracket (also called BFV Poisson bracket).

More precisely, for each field $z^\alpha$ one introduces the
``momenta'' $\pi_\alpha$ with $\p{\pi_\alpha}=\p{z^\alpha}$ and
$\gh{\pi_\alpha}=-\gh{z^\alpha}$ in the graded symmetric case and the
``antifields'' $z^*_\alpha$ with $\p{z^*_\alpha}=\p{z^\alpha}+1$ and
$\gh{z^*_\alpha}=-\gh{z^\alpha}+1$, with the natural extensions for
the derivatives of momenta and antifields. The horizontal complex is
then extended to include either the momenta or the antifields and
their derivatives. We introduce a subscript $E$ to denote elements of
the extended complex. A graded symmetric (skew-symmetric) functional
$k$-vector is then a local functional of homogeneity $k$ in the
momenta (antifields) and their derivative. There is a map from
functional multivectors $\int \omega^{g,n}_E=\int d^nx\, f^{g}_E $ to
evolutionary vector fields on the extended complex defined
through\footnote{We write down the formulas explicitly only for the
  symmetric case.  The skew-symmetric case can be obtained by
  substituting $\pi_\alpha$ with $z^*_\alpha$ and changing the
    sign-factors appropriately.}
\begin{gather}
  \label{eq:9}
  \left\{\int d^nx\, f^{g}_E ,\cdot \right\}_E=-\d_{(\mu)}
  \vddr{f^{g}_E}{\pi_\alpha}\dover{^S}{z^{\alpha}_{(\mu)}}+(-1)^{\p{\alpha}}(\pi_\alpha
  \rightleftarrows z^\alpha) \,.
\end{gather}
Using multiple integrations by parts and \eqref{eq:26}, it is then
easy to see that this map induces a well defined even (odd)
graded Lie bracket in the space of functional multivectors. 

The BRST differential $s$ itself is then the evolutionary vector field
generated by the univector
\begin{gather}
  \label{eq:20}
  \Omega_0=-\int d^nx\, S^\alpha \pi_\alpha,\quad \half \big\{
  \Omega_0,\Omega_0\big\}_E=0,\quad\gh{\Omega_0}=1\,. 
\end{gather}

The action of the BRST differential in the space of functional
multivectors is then simply the adjoint action of $\Omega_0$
\begin{gather}
  s_E \int \omega^{g,n}_E=\pb{\Omega_0}{\int\omega^{g,n}_E}_E\,.
\end{gather}
In the space of functional univectors, this action is isomorphic to
the commutator action of $s$ in the space of evolutionary vector
fields.

Given a functional $k$-vector represented by a local functional $V_E$
it determines a well defined graded-symmetric $k$-multilinear
operation on the space of local functionals of the original
(non-extended) complex in a standard way.  This can be expressed using
the so-called derived bracket\footnote{See e.g.  \cite{ThVoronov} for
  more details on derived brackets.}: if $F_1,\ldots, F_k$ are local
functionals of the non-extended complex, identified as
$\pi$-independent functionals of the extended complex, then
\begin{equation}
  V(F_1,\ldots,F_k)=\frac{1}{k!}\pb{\ldots\pb{\pb{V_E}{F_1}_E}{F_2}_E}{\ldots F_k}_E\,.
\end{equation}
This operation is well-defined on local functionals and gives a local
functional of the non-extended complex as can be easily seen by
counting homogeneity in $\pi$.

For instance, a functional bivector $\Omega_1$ of unit parity and unit
ghost number satisfying $\half\pb{\Omega_1}{\Omega_1}_E=0$ corresponds to a
functional antibracket.  The cocycle condition $s_E\Omega_1=0$ then
means that the bracket is $s$-invariant, i.e., that $s$ differentiates
the antibracket.

\section{Local BRST cohomology for AKSZ-type sigma models}
\label{sec:appl-aksz-type}

\subsection{Cohomology of space-time part}
\label{sec:ident-ladd-fields}

{\em The cohomology of the space-time part of the BRST differential
  for AKSZ-type sigma models is derived by using results available in
  the literature.}

The coefficients of $\Psi^A(x,\theta)$ in an expansion as series in
$\theta^\mu$ constitute the field/antifield content of AKSZ-type sigma
models,
\begin{equation}
  \label{eq:18a}
z^\alpha\equiv(
\Psi^A,\Psi^A_\mu,\dots,\Psi^A_{\mu_1\dots\mu_k},\dots,\Psi^A_{\mu_1\dots\mu_n})\,.
\end{equation}
The $z^\alpha$ thus consist of ``formfields'', a set of
fields/antifields which are completely skew-symmetric\footnote{We use
  round (square) brackets to denote normalized
  (skew)-symmetrization. }  in the spacetime indices,
$\Psi^A_{\mu_1\dots\mu_k}=\Psi^A_{[\mu_1\dots\mu_k]}$ and contain all
possible form degrees, $k=0,\dots,n$.  Furthermore, we assign
$\gh{\Psi^A_{\mu_1\dots\mu_k}}=\gh{ \Psi^A}-k$.  In the jet-space
context, we can introduce an object analogous to the map $\Psi^A(x,\theta)$,
the ``complete ladder fields'' in the terminology of
\cite{Carvalho:1995uu,Piguet:1995er},
\begin{equation}
  \label{eq:19a}
  \tilde \Psi^A=\sum_{k=0}^n\Psi_k^A,\quad \Psi^A_k=
\frac{1}{k!}\Psi^A_{\mu_1\dots\mu_k}\theta^{\mu_1}
  \dots\theta^{\mu_k}\,,
\end{equation}
where $\Psi^A_0\equiv\Psi^A$. 

We refer to the first term in the BRST differential \eqref{eq:s-AKSZ}
involving the de Rham differential $\derham$ as the spacetime part and
denote it by $s_{-1}$. When translated~\footnote{To simplify notations in
  this section, we redefine the BRST differential by an overall factor
  $(-1)^{n}$ and change the sign of the term in $s$ involving
  $\derham$. This can be achieved by the transformation
  $\theta^\mu\to-\theta^\mu$.} in the jet-space context, we have
$s_{-1}\tilde\Psi^A=-d_H\tilde\Psi^A$, or, more explicitly,
\begin{equation}
  \label{eq:20a}
  s_{-1}\Psi^A=0,\quad  s_{-1}\Psi^A_{\mu_1\dots\mu_k}=-(-)^{A+k-1}
k\,\d_{[\mu_1}\Psi^A_{\mu_2\dots\mu_k]},
\end{equation}
which can be summarized by 
\begin{equation}
  \label{eq:33}
  \tilde{s_{-1}} \tilde \Psi^{ A}=0\,,\qquad \tilde{s_{-1}}=s_{-1}+d_H\,.
\end{equation}

As discussed in detail in the proof of theorem 3.1 of
\cite{Henneaux:1998rp}, the idea is to decompose the form fields and
their derivatives $\d_{\nu_1}\dots\d_{\nu_m}\Psi^A_{\mu_1\dots\mu_k}$
into irreducible tensors under the general linear group $GL(n)$.  One
then finds that all the field variables form contractible pairs except
for the undifferentiated $\Psi^A$. Compared to the situation
considered in \cite{Henneaux:1998rp} no curvatures remain because the
last formfield $\Psi^{ A}_{\mu_1\dots\mu_n}$ is of maximal degree $n$.  The
cohomology $H(s_{-1},\hat\Omega)$ can thus be described by functions of
$\Psi^{ A},x^\mu,\theta^\mu$ alone,
\begin{eqnarray}
  \label{eq:12}
  H(s_{-1},\hat\Omega)\simeq\{\lambda(x,\theta,\Psi^{ A})\},
\end{eqnarray}
where $\lambda(x,\theta,\Psi^{ A})$ contains no field independent terms,
$\lambda(x,\theta,0)=0$. 

The analysis of the descent equations is then standard. In the present
case, it is presented in
\cite{Carvalho:1995uu,Piguet:1995er,Henneaux:1998rp} (see also section
14 of \cite{Barnich:2000zw}).  Using the ordinary Poincar\'e lemma on
the base space, one finds
\begin{eqnarray}
H(\tilde{s_{-1}},\hat\Omega)
\simeq\{\nu(\tilde \Psi^{ A})\}\label{eq:7} \,,
\end{eqnarray}
where $\nu$ denotes a polynomial in its arguments without constant
term. This
cohomology is isomorphic to the cohomology of $s_{-1}$ in the space $\hat
\Omega^{*,0}_{x=0}$ of $x$-independent zero forms
\begin{equation}
  \label{eq:4}
  H(\tilde{s_{-1}},\hat\Omega)\simeq H(s_{-1},\hat
\Omega^{*,0}_{x=0}),
\end{equation}
since $H(s_{-1},\hat \Omega^{*,0}_{x=0})\simeq\{\nu(\Psi^{ A})\}$.
Finally, 
\begin{equation}
H(s_{-1},\hat\cF)\simeq\{\nu(\Psi^{ A})|_n\}\label{eq:37}\,,
\end{equation}
where $|_n$ means that one should restrict oneself to the form
of top degree $n$ in an expansion according to form degree. 

\subsection{Cohomology of complete differential}
\label{sec:reduct-brst-cohom}

{\em We show that the field theoretic BRST cohomology in the space of
  local functionals is isomorphic to $Q$-cohomology of target space
  functions in coordinate neighborhoods.}

The full BRST differential for AKSZ-type sigma models is defined by 
\begin{equation}
  \label{eq:31}
  \tilde s \tilde \Psi^A= Q^A(\tilde\Psi)\,,
\end{equation}
and is extended by prolongation to the derivatives of the formfields.

Consider then the degree which counts the number of form indices minus
the number of $\theta$'s, 
\begin{equation}
  \label{eq:36}
  \cN=\sum_{k=0}^n
  k\d_{(\lambda)}\Psi^A_{\mu_1\dots\mu_k}\dover{}{\Psi^A_{(\lambda)
      \mu_1\dots\mu_k}}-\theta^\mu\dover{}{\theta^\mu}. 
\end{equation}
According to this degree, the space of horizontal forms is bounded
from below. Furthermore, $s=s_{-1}+s_0$, where $s_{-1}$ is the
space-time part discussed previously.  In particular, in
form degree zero, one finds that
\begin{equation}
  \label{eq:28}
  s_0  \Psi^A= Q^A(\Psi)\,,
\end{equation}
and is thus entirely determined through the homological vector field
$Q$ in the target space $\manM$. The action of
$s_0$ on the remaining fields contained in $\Psi^A_k,
k\geq 1$ is then determined by the same $Q^A$ by expanding
\eqref{eq:31} according to higher form degrees, $s_0
\Psi^A_k=Q^A(\tilde\Psi)\big|_k$ and taking into account
\eqref{eq:33}. Explicitly
\begin{gather}
  \label{eq:32}
 s_0 \Psi^A_1=\Psi^B_1\dover{Q^A}{\Psi^B}(\Psi),\\
 s_0 \Psi^A_2=\Psi^B_2\dover{Q^A}{\Psi^B}(\Psi)+\half
 \Psi^{B_1}_1\Psi^{B_2}_1 \dover{ Q^A}{\Psi^{B_2}\d\Psi^{B_1}}(\Psi),\\
\vdots\nonumber
\end{gather}

The cohomology $H^{g+n}(\tilde s,\hat \Omega)$ reduces to the
cohomology of $s_0$ induced in $H^{g+n}(\tilde{s_{-1}},\hat\Omega)$
because $\tilde s=\tilde{s_{-1}}+s_0$. Note that
$s_0\tilde\Psi^A=Q^A(\tilde\Psi)$ on account of \eqref{eq:33} and
\eqref{eq:31} so that $H(\tilde s,\hat \Omega)\simeq
H(s_0,\nu(\tilde\Psi^A))$.  Since $H(s,\cF)\simeq H(\tilde
s,\hat\Omega)$ on the one hand and $H(s_0,\nu(\tilde\Psi^A))\simeq
H(Q,\nu(\Psi^A))$ on the other, we have thus shown 
\begin{prop}\label{prop1}
  The local BRST cohomology $H(s,\hat\cF)$ is isomorphic to
  the cohomology $H(Q)$ in target space functions for coordinate
  neighborhoods of the base and the target space,
\begin{equation}
  \label{eq:3}
  H^g(s,\hat \cF)\simeq H^{g+n}(Q)\,,
\end{equation}
where
\begin{equation}
  \label{eq:16}
H^g(Q)\ni  [\Theta^g_{\alpha_g}(\Psi^A)]\longleftrightarrow
[\Theta^g_{\alpha_g}(\tilde \Psi^A)|_n] \in H^{g-n}(s,\hat \cF),
\end{equation}
with $[\Theta^g_{\alpha_g}(\Psi^A)]$ denoting representatives of
$H^g(Q)$.
\end{prop}

{\bf Remarks:} 

(i) By identifying integrals of functions evaluated for maps of
compact support with the algebraic version of local functionals, one
can consider the map $\cI$ defined in \eqref{fF-map} as a map from
functions on target space to local functionals.  The proposition can
then be reformulated by the statement that $\cI$, for AKSZ-type sigma
models, is locally an isomorphism in cohomology or, more precisely, a
quasi-isomorphism of complexes in the case without bracket and a
quasi-isomorphism of differential graded Lie algebras in the case with
bracket.

(ii) In the case of the $1$-dimensional AKSZ sigma models associated
with Hamiltonian BFV systems with vanishing Hamiltonian,
Proposition~\bref{prop1} states that the Poisson algebra of
Hamiltonian BRST cohomology and the antibracket algebra of Lagrangian
BV cohomology in the space of local functionals are locally isomorphic, as
originally derived in \cite{Barnich:1996mr}.

(iii) The BRST extension
\cite{Barnich:2004cr,Barnich:2005ru,Barnich:2006pc} of the unfolded
linear equations
\cite{Vasiliev:1988xc,Vasiliev:1988sa,Vasiliev:1994gr,Vasiliev:2005zu}
developed originally in the context of higher spin gauge fields is
almost of the above form. Indeed, in this case, there are also only
complete ladder fields but instead of (\ref{eq:31}), one has more
generally
\begin{equation}
\tilde s\tilde
\Psi^A=Q^A(x^\mu,\theta^\mu,\Psi^B_{\mu_1,\dots\mu_k})\label{eq:17}.
\end{equation}
In some cases, this BRST differential can be seen as the linearization
of some nonlinear AKSZ differential around a particular solution that
brings the explicit dependence on $x,\theta$~\cite{Grigoriev:2006tt}.
It would be interesting to compute the BRST cohomology in the space of
local functionals for this more general case along these lines. We
plan to return to this question elsewhere.

(iv) Although the base space of the AKSZ sigma model is a
supermanifold, we have used the standard jet-space technique and
considered the component fields $\Psi^A_{\mu_1\ldots \mu_k}$ instead
of using superfields. An elegant alternative would be to work directly
with the extension of \cite{Khudaverdian:2001qe} to jets on
super-base spaces.

\subsection{Cohomology for functional multivectors}

{\em The isomorphism of the field theoretic cohomology and the
  cohomology for target space functions is extended to the case of the
  BRST differential acting in the space of functional mulitvectors by
  showing that the latter is again of AKSZ-type.}
 
In order to discuss the cohomology in the space of graded symmetric
(skew-symmetric) functional multivectors, one introduces the
conjugate momenta $\pi_{A}^{\mu_1\dots\mu_k}$ (antifields $
\Psi_{A}^{*\mu_1\dots\mu_k}$) and considers the functional
\begin{equation}
  \label{eq:5}
  \Omega_0=-\int d^nx\, \sum_{k=0}^n
 \frac{1}{k!} s\Psi^A_{\mu_1\dots\mu_k}\pi_A^{\mu_1\dots\mu_k}\,.
\end{equation}
In the AKSZ setting, it is then natural to combine the momenta
(antifields) into superfields $\tilde \Pi_{A}$ in such a way
that \eqref{eq:5} takes the form
\begin{equation}
  \label{eq:6}
  \Omega_0=-\int d^nxd^n\theta\,  s\tilde \Psi^A\tilde \Pi_A
  =-\int d^nxd^n\theta\,  \big[-d_H\tilde \Psi^A\tilde
  \Pi_A+Q^A(\tilde\Psi)\tilde \Pi_A\big]\,.
\end{equation}
Explicitly, one has
\begin{equation}
   \label{eq:34}
  \tilde \Pi_A=\sum_{k=0}^n\Pi^{n-k}_A,\quad
  \Pi^{n-k}_A=\frac{(-)^{n+k(|A|+1)}}{k!(n-k)!}
  \pi_{A}^{\mu_1\dots\mu_k}\epsilon_{\mu_1\dots\mu_k\nu_{k+1}\dots\nu_n}
\theta^{\nu_{k+1}}\dots\theta^{\nu_n}\,,
\end{equation}
where $|A|$ is a shortcut for $|\Psi^A|$.  For the adjoint action of
$\Omega_0$ one then finds
\begin{equation}
  \label{eq:8}
s\tilde \Psi=-d_H \tilde\Psi+Q^A(\tilde\Psi)\,, \qquad 
  s\tilde\Pi_A=-d_H\tilde\Pi_A-(-1)^{\p{A}}
\ddl{Q^B}{\Psi^A}(\tilde\Psi)\tilde\Pi_B\,.
\end{equation}

The BRST differential \eqref{eq:8} is then again of AKSZ-type. The
associated target space is given by the (odd) cotangent bundle
$(\Pi)T^* \manM$, with canonical (odd) Poisson structure
\begin{equation}
  \label{eq:15}
  \pb{\Pi_B}{\Psi^A}_{(\Pi)T^*\manM}=-\delta_{B}^A\,,\quad
  \gh{\Pi_A}=-\gh{\Psi^A}+n\,, 
\end{equation}
and homological vector field
\begin{equation}
  Q_E=\pb{-Q^A\Pi_A}{\cdot}_{(\Pi)T^*\manM}=Q^A\ddl{}{\Psi^A}-
(-1)^{\p{A}}\ddl{Q^B}{\Psi^A}\Pi_B\ddl{}{\Pi_A}\,.
\end{equation}
Furthermore, in terms of the map $\Pi_A(x,\theta)$, the canonical Poisson
bracket can be identified with a Poisson bracket of the form
\eqref{eq:path-bracket}.  

It then follows from Proposition~\bref{prop1} that the BRST cohomology
in the space of functional multivectors is locally isomorphic to the
cohomology of $Q_E$, or, more precisely, that the map $\cI_E$, sending
the target space multivectors to functional multivectors according to
\begin{equation}
\cI_E {\,\rm:\,} f_E \quad \mapsto \quad \int_{\manX} d^nxd^n\theta f_E\,,
\end{equation}
where $f_E\in \cA_E,$ with $\cA_E$ denoting the space of functions on
the target space $(\Pi)T^*\manM$, is locally a quasi-isomorphism of
differential graded Lie algebras.

Finally, as an illustration, we note that
expression~\eqref{eq:path-bracket} of the bracket on the space of maps
can be interpreted as an explicit
realization of the map $\cI_E$ for (skew)-symmetric $2$-vectors, when
identifying algebraic local functionals with integrals of target space
functions evaluated for maps of compact support.

\section{Functional multivectors, symmetries and generalized Poisson
  structures}
\label{sec:applications}

\subsection{Applications of BRST cohomology for functional
  multivectors}
\label{sec:appl-brst-cohom}

{\em It is pointed out that BRST cohomology in the space of functional
  multivectors is relevant for classifying symmetries of the equations
  of motion and weak Poisson or Lagrange structures.}

Consider a gauge theory for which a Lagrangian does not exist or is
not (yet) specified. Such a theory can still be described in terms of
a BRST differential $s$ that is not necessarily generated by a master
action in an appropriate antibracket. As pointed out in ~\cite{\BGST},
consistent deformations of such theories are then controlled by the
adjoint cohomology of $s$ in the space of evolutionary vector fields,
or in other words, by $H^1_1(s_E,\cF_E)$.

The identification of the cohomology groups controlling global
symmetries requires some care.  To fix ideas, let us first consider
the case of partial differential equations of motion of
Cauchy-Kovalevskaya type, as considered for instance in
\cite{Dickey:1991xa,Olver:1993}, for which $s$ reduces to the
so-called Koszul resolution
\cite{Fisch:1989dq,Fisch:1990rp,Henneaux:1991rx} of the surface
defined by the equations in jet-space. In this case, it is
straightforward to check that $H^0_1(s_E,\cF_E)$ coincides with the
usual definition of equivalence classes of symmetries of the equations
of motion, i.e., evolutionary vector fields that leave the equations
of motion invariant quotiented by such vector fields that vanish when
the equations of motion hold.

In the case of variational equations of motions, with non trivial
relations between the equations and their derivatives (``Noether
identities''), there is a well defined concept of a proper solution to
the BV master equation. In this case, $H^0_1(s_E,\cF_E)$ is given by
equivalence classes of equations of motion symmetries quotiented not
only by the ones vanishing when the equations of motion hold, but in
addition by all non trivial gauge symmetries (see
e.g.~\cite{Barnich:2000zw}).

In the non variational case, $H^0_1(s_E,\cF_E)$ is again given by the
quotient space of equations of motion symmetries modulo evolutionary
vector fields that vanish when the equations of motion hold and modulo
the non trivial gauge symmetries encoded in $s$. The question is then
whether the latter include all the non trivial gauge symmetries, which
in turn hinges on an appropriate non-variational version of properness
for the BRST differential $s$. The precise definition of this concept
is beyond the scope of the present work\footnote{In the
  finite-dimensional setting an appropriate notion of properness was
  proposed in~\cite{Kazinski:2005eb}.  Its generalization to the local
  field theory setting is not entirely straightforward.}. In what
follows we simply assume that all the gauge symmetries and
reducibility relations between the equations and between the gauge
generators are accounted for by the BRST differential.

Given a BRST differential $s$, another natural question is whether the
gauge theory determined by $s$ admits a Lagrangian or Hamiltonian
description. In the case of non-gauge systems, this question is known
as the inverse problem of the calculus of variations. In the BRST
context, it translates into the question of existence of a generator
for $s$ in an appropriate bracket that is usually assumed non
degenerate. One can distinguish two cases. In the Lagrangian or BV
case, the bracket\footnote{In the space of local functionals, these
  brackets are of the Gelfand-Dickey-Dorfman type
  \cite{Gelfand:1979zh},\cite{Dickey:1991xa}, see also
  \cite{Barnich:1996mr} for the current context.} for the fields
$z^\alpha$ is an odd graded bracket of ghost number $1$, while the
canonical generator, the solution of the BV master equation, is even
of ghost number $0$. In the Hamiltonian BFV case, the Poisson bracket
is even of ghost number $0$, while the canonical generator, the BRST
charge, is odd of ghost number $1$.  When $s$ is proper in the sense
discussed above, this question is equivalent to the question whether
the equations it encodes are variational in the Lagrangian case or
whether the constraints it describes are first class
(``co-isotropic'') in the Hamiltonian case.

Let us for definiteness restrict ourselves to the case of an odd
bracket on the space of fields and hence to the Lagrangian BV
picture. The question of existence of a Lagrangian for a gauge theory
can be adressed using the notion of Lagrange
structure~\cite{Kazinski:2005eb},
which is the Lagrangian counterpart of a possibly weak and degenerate
Poisson structure of the Hamiltonian formalism~\cite{Lyakhovich:2004xd}.
In the BRST theory terms the Lagrange structure can be
represented~\cite{Kazinski:2005eb} as a strong homotopy Schouten algebra
structure (see e.g.~\cite{ThVoronov,KV}).

In local field theory, such a structure can be defined as a collection
of $n$-ary functional multivectors satisfying appropriate
compatibility conditions including, in particular, the Jacobi identity
for the bracket induced in $s_E$-cohomology. More concretely, it can
be defined as a deformation of \eqref{eq:20} by terms of higher order
in $\pi_\alpha$, $\Omega=\Omega_0+\Omega_1+\Omega_2+\ldots$ with
$\gh{\Omega}=1$, where $\Omega_k$ denotes a local functional that is
homogeneous of degree $k+1$ in $\pi_\alpha$ and their derivatives. As
usual, the compatibility conditions are combined into the master
equation\footnote{This is the local field theory version of the master
  equation considered in~\cite{Kazinski:2005eb,Lyakhovich:2004xd}.
  Master equations of this type have been also considered
  in~\cite{Batalin:1992} from a different perspective.}

\begin{equation}
\label{me}
\half \pb{\Omega}{\Omega}_E=0 \Longleftrightarrow\quad \left\{\begin{array}{c}
 s_E \,{\Omega_1}=0\,, \\ \half\pb{\Omega_1}{\Omega_1}_E+ s_E \,
{\Omega_2}=0\,,\\ \pb{\Omega_1}{\Omega_2}_E+ s_E\,
{\Omega_3}=0\,,\\ \vdots
\end{array}
\right.
\end{equation}
Two such deformations $\Omega$ and $\Omega^\prime$ are considered
equivalent if there exists a local functional $\Xi=\sum_{k\geq
  1} \Xi_k$ such that $\Omega^\prime=\exp{\pb{\Omega_0}{\cdot}}_E\,\Xi$,
where $\Xi_k$ is homogeneous of degree $k+1$ in $\pi_\alpha$.

In particular, non trivial first order Lagrange structures are controlled
by $H^1_2(s_E,\hat \cF_{E})$, the cohomology of $s_E$ in the space of
functional bivectors of ghost number $1$, while the second equation on
the right of \eqref{me} encodes the Jacobi identity satisfied in BRST
cohomology. 

In the standard deformation approach for gauge theories
\cite{Barnich:1993vg,Henneaux:1997bm,Stasheff:1997fe}, it is crucial
to take due care of locality since otherwise the deformation theory is
trivial in the sense that all first order deformations extend to
complete deformations. This is also true for Lagrange or weak and
degenerate Poisson structures. More precisely, the classification
result in \cite{Kazinski:2005eb} stating that all the Lagrange
structures are trivial in the finite dimensional case will not
generally hold once field theoretic locality is taken into account.
Indeed, examples of nontrivial Lagrange structures for field theories
were provided in~\cite{Kazinski:2005eb,Lyakhovich:2007cw}.

As defined above, a Lagrange structure is an equivalence class
$[\Omega]$ of deformations of $\Omega_0$ in the space of functional
multivectors. This is consistent with the point of view adopted
in~\cite{\BGST} that being Lagrangian or not is a property of
equivalence classes of equations of motion under addition/elimination
of generalized auxiliary fields, because generalized auxiliary fields
correspond to contractible pairs for $s_E$. 

\subsection{Consequences for AKSZ-type sigma models}
\label{sec:cons-aksz-type}

{\em As a direct consequence of the main result, the classification of
  Lagrange or weak Poisson structures simplifies to a
  target space problem for AKSZ-type sigma models.}

As we have seen, the field theoretic BRST cohomology of AKSZ-type
sigma models originates from $Q$-cohomology of target space
functions, both for standard local functionals and for functional
multivectors. It then follows from standard deformation theory
arguments that Lagrange or weak Poisson structures for these models
can be entirely discussed in the target space, or in other words, that
one can consistently get rid of the space-time derivatives and of the
higher forms in the Lagrange/Poisson structure of these models.

Indeed, for $\Omega$ satisfying \eqref{me}, the term $\Omega_1$
quadratic in $\pi_\alpha$ and their derivatives can for instance be
written as $\Omega_1=\cI_E(\omega_1)+s_E \Xi_1$ for some $\omega_1\in
\cA_{E}$ and $\Xi_1\in \hat\cF_E$. By exponentiating the
transformation generated by $\Xi_1$ one arrives at an equivalent
$\Omega$ with the same $\Omega_0$ but $\Omega_1=\cI_E(\omega_1)$. At
the next order, one finds $s_E\Omega_2+\half \cI_E
\pb{\omega_1}{\omega_1}_{(\Pi)T^*\manM}=0$. A standard reasoning
involving contractible pairs implies that
$\Omega_2=\cI_E(\omega_2)+s_E\Xi_2$ with $Q_E \omega_2+\half
\pb{\omega_1}{\omega_1}_{(\Pi)T^*\manM}=0$ and again, through
exponentiation, one arrives at an equivalent $\Omega$ such that
$\Omega_2=\cI(\omega_2)$. Going on in this way for the higher orders,
one ends up with an equivalent $\Omega$ of the form
\begin{equation}
  \Omega=\Omega_0+\cI_E(\omega)=\Omega_0+
  \int_{\manX} d^nxd^n\theta\, (\omega_1+\omega_2+\ldots) \,.
\end{equation}
Here $\omega_k\in \cA_{E}$ is a polynomial of order $k+1$ in the
momenta $\pi_A$.  It follows that $\omega=-Q^A\pi_A+\sum_{k\geq1}
\omega_k$ satisfies the target space master equation 
\begin{equation}
  \half\pb{\omega}{\omega}_{(\Pi)T^*\manM}=0\,.
\end{equation}

Finally, for the sake of illustration, let us consider Chern-Simons
theory based on a simple Lie algebra. In this case, the target space
of the extended model is the $(n|n)$-dimensional supermanifold $M=\Pi
T^* \cG$ with coordinates $c^a,\,\gh{c^a}=1$ and
$\pi_a,\,\,\gh{\pi_a}=2$ and QP structure determined by
\begin{equation}
 \pb{\pi_a}{c^b}_{M}=-\delta_a^b\,,
\qquad Q_E=-\pb{\half c^a c^b f_{ab}^c\pi_c}{\cdot}_M\,.
\end{equation}
The cohomology of $Q_E$ can be identified with the Lie algebra
cohomology of $\cG$ with coefficients in polynomials in $\pi_a$.

For a simple Lie algebra $\cG$, this cohomology is given by the
algebra generated by the primitive elements, which are at least cubic
in $c^a$, tensored with the invariant polynomials in $\pi_a$.  The
cohomology in the space of elements linear in $\pi_a$ is empty and one
concludes that the Chern-Simons theory is rigid and does not have
nontrivial symmetries at the level of the equations of motion.

The cohomology in the space of elements quadratic in $\pi_a$ is given
by the invariants $g^{ab}\pi_a\pi_b$, with $g^{ab}$ the inverse of the
Killing form, tensored with the algebra of primitive elements. First
order Lagrange structures are classified by ghost number one local
functionals of the extended model that are quadratic in the
$\pi$. Using the isomorphism, these can be represented by ghost number
$4$ target space functions quadratic in $\pi_a$, i.e., by
$g^{ab}\pi_a\pi_b$. There is thus only one non trivial cohomology
class and hence only one nontrivial first order Lagrange structure. It
trivially extends to higher orders and obviously coincides with the
usual one, i.e., the standard BV antibracket of the AKSZ description.

\section*{Acknowledgements}
\label{sec:acknowledgements}

\addcontentsline{toc}{section}{Acknowledgments}

The authors thank X.~Bekaert and F.~Brandt for useful
discussions. This work has been initiated during a stay at the Erwin
Schr\"odinger International Institute for Mathematical Physics and is
supported in parts by the Fund for Scientific Research-FNRS (Belgium),
the International Solvay Institutes, by the Belgian Federal Science
Policy Office through the Interuniversity Attraction Pole P6/11 and by
IISN-Belgium. The work of M.G.  is supported by the RFBR grant
08-01-00737, RFBR-CNRS grant 09-01-93105.


\def\cprime{$'$}
\providecommand{\href}[2]{#2}\begingroup\raggedright\endgroup

\end{document}